\documentclass[runningheads]{llncs}
\usepackage[T1]{fontenc}
\usepackage{graphicx}
\usepackage{booktabs}
\usepackage[misc]{ifsym}

\newcommand{\corr}{(\Letter)}
\usepackage{multirow}  %
\usepackage{booktabs}  %
\usepackage{latexsym}
\usepackage{amssymb}
\usepackage{amsmath}
\usepackage[hidelinks]{hyperref}
\usepackage{booktabs}
\usepackage{enumitem}
\usepackage{graphicx}
\usepackage{color}
\usepackage{appendix}
\usepackage{comment}
\usepackage[space]{cite}
\usepackage{tikz}
\usetikzlibrary{shapes, arrows, positioning, fit, shapes.geometric}
\usepackage{caption}
\usepackage{wrapfig}

\usepackage{amssymb}%
\usepackage{pifont}%

\usepackage{cleveref}
\crefformat{section}{#2Sec.~#1#3}
\crefformat{figure}{#2Fig.~#1#3}
\crefformat{algorithm}{#2Alg.~#1#3}
\crefformat{table}{#2Tab.~#1#3}
\crefformat{equation}{#2Eq.~#1#3}
\crefformat{definition}{#2Def.~#1#3}
\crefformat{appendix}{#2App.~#1#3}
\crefmultiformat{equation}{Equations #2#1#3}{ and~#2#1#3}{, #2#1#3}{ and~#2#1#3}
\crefmultiformat{table}{Tables #2#1#3}{ and~#2#1#3}{, #2#1#3}{ and~#2#1#3}

\usepackage{todonotes} %
\newcommand{\nardine}[1]{}
\newcommand{\robin}[1]{}

\usepackage{subcaption}

\begin{document}

\title{Game of Phones: Harnessing Game Theory and LLMs in `Bot Wars' to Counteract Phone Scams}

\titlerunning{Game of Phones}

\author{Nardine Basta \corr \and
Robin Carpentier \and
Benjamin Zi Hao Zhao \and Dali Kaafar}
\institute{Macquarie University, NSW, Australia\\
\email{nardine.basta@mq.edu.au}}
 \tocauthor{Nardine Basta, Robin Carpentier, Benjamin Zi Hao Zhao, Dali Kaafar}  
 \toctitle{Game of Phones: Harnessing Game Theory and LLMs in `Bot Wars' to Counteract Phone Scams} 
\authorrunning{N. Basta et al.}

\maketitle              %

\begin{abstract}
Phone scams impose substantial and growing costs on individuals and financial systems worldwide, yet existing countermeasures impose no resource cost on perpetrators. Strategic scam-baiting, in which automated agents engage scammers to deplete their operational capacity, offers a proactive alternative but requires dynamic strategy adaptation that static systems cannot provide. We introduce a game-theoretic framework enabling LLMs to learn effective counter-strategies through in-context experience without fine-tuning. The framework models scammer-baiter interactions as a two-player game with empirically grounded strategies and utility functions, using Nash equilibrium as a formal benchmark for evaluating learned behaviour. Across 300 simulated dialogues and three LLM architectures, adaptive strategy selection achieves 28.9--79.7\% closer alignment to Nash equilibrium than non-adaptive baselines.%
The code and an annotated dataset of 300 dialogues with turn-level strategy labels and utility scores are released\footnote{Code and dataset are available at \href{https://github.com/nardinebasta/GameofPhones_}{github.com/nardinebasta/GameofPhones\_/}\label{footnoteCodeRepo}}. 
\end{abstract}

\section{Introduction}
Phone scams impose substantial and growing costs on individuals and financial systems worldwide. In 2024, the U.S. Federal Trade Commission recorded consumer fraud losses exceeding \$12 billion, with phone calls remaining one of the dominant delivery vectors~\cite{FTC2024}. %
The increasing integration of generative AI into scam infrastructure has further accelerated the pace at which scripts evolve, significantly outpacing the adaptation cycles of conventional defences~\cite{wood2023analysis}.

Existing countermeasures like call blocking, blacklisting, and ML-based detection~\cite{pandit2018measuring, derakhshan2021Detecting} operate reactively. They impose no cost on the scammer, who can immediately redirect resources towards new targets. \textit{Scam-baiting}, in which an agent impersonates a potential victim to waste scammers' time, offers a fundamentally different posture. Early automated approaches~\cite{LLM2, basta2025botwars} demonstrate feasibility but rely on static templates and fixed heuristics, leaving them ineffective against operators who adapt to recognise and disengage from bots.

Effective automated baiting requires resolving three main challenges: (i) dynamically balancing the conflicting objectives of prolonging the interaction and maintaining credibility; (ii) employing baiting strategies that are effective relative to the scammer's observed behaviour; and (iii) evaluating agent adaptation through formal metrics beyond interaction length alone. We address these challenges through a game-theoretic framework enabling LLMs to learn effective baiting strategies via in-context learning, without parameter updates. 

The framework operates after a call has been flagged as a scam and models scammer-baiter interactions as a two-player non-cooperative game. Strategies employed by each player are grounded in documented scam techniques~\cite{wood2023analysis, abagnale2019scam} and persuasion research~\cite{cialdini1984influence}, and payoffs are attributed to players based on their performance relative to their respective objectives. To build an optimal scam-baiter --- one that consistently selects strategies that maximise scammer time wastage --- Nash equilibrium (NE) is computed from empirically constructed payoff matrices. A baiter whose strategy distribution aligns with NE maximises its guaranteed performance against any scammer behaviour, making equilibrium proximity a principled measure of how effectively an agent has learned to play, and serving as the benchmark against which learned LLM behaviour is evaluated across repeated games. %

This paper addresses the following research questions: \textbf{RQ1} Can LLMs learn effective baiting strategies through in-context learning from repeated interactions, without fine-tuning? \textbf{RQ2} Do LLMs converge towards NE strategy distributions through experience-based adaptation, and does proximity to equilibrium correlate with baiting effectiveness? \textbf{RQ3} Which baiting strategies are most effective across LLM architectures, and to what extent do optimal strategies vary with architectural differences? \textbf{RQ4} Which LLM architecture offers the most favourable trade-off between baiting effectiveness and computational cost for production deployment?%

We evaluate the framework through 300 simulated dialogues across DeepSeek, Mixtral, and GPT-4. Delay and Counter Questioning emerge as universally effective baiting tactics, while optimal mixing ratios are architecture-dependent. Adaptive strategy selection achieves 28.9--79.7\% closer alignment to NE than non-adaptive baselines. DeepSeek achieves the strongest baiting performance ($\approx60\%$ baiter win rate; 42-turn average duration) at substantially lower cost than GPT-4, offering a favourable cost-performance trade-off for deployment. %

The paper makes four contributions: (1) Empirically validated identification of effective baiting strategies via NE analysis, with architecture-specific deployment guidance.
(2) An empirical demonstration that LLMs acquire effective counter-strategies through in-context learning without fine-tuning, achieving measurably closer alignment to the Nash equilibrium than non-adaptive baselines across all models tested. (3) A game-theoretic evaluation framework for adaptive strategy learning in adversarial human-agent dialogues, providing the first formal assessment mechanism for LLM adaptation in scam-baiting contexts. (4) An annotated dataset of 300 dialogues with turn-level strategy labels and utility scores, released\textsuperscript{\ref{footnoteCodeRepo}} to support reproducibility and future research.

\section{Related Work}
\textbf{Scam Detection and Operational Defences.} Deployed countermeasures against phone scams span network-level filtering (call blocking, telephony blacklisting~\cite{pandit2018measuring}), content-based detection (ML classifiers~\cite{li_machine_2018}, NLP script detection~\cite{derakhshan2021Detecting}), and caller authentication with virtual assistant mediation~\cite{usenix1}. All operate reactively, and caller ID spoofing further undermines filtering reliability~\cite{usenix3}. Critically, no deployed system imposes a direct resource cost on scam operators; a blocked or detected scammer can immediately redirect effort towards a new target at negligible cost. This asymmetry motivates the resource-depletion approach pursued in this work.

\textbf{Automated Scam-Baiting.} Manual scam-baiting has been studied as a deliberate countermeasure~\cite{dynel_you_2021}, with documented evidence that sustained engagement meaningfully occupies scammer resources. Automation efforts remain limited. Bajaj and Edwards~\cite{LLM2} demonstrated LLM-generated victim responses but relied on static templates without strategy adaptation. Basta et al.~\cite{basta2025botwars} introduced multi-agent interaction and persona diversity, yet provided no formal strategic modelling, performance-based adaptation, or quantitative evaluation metrics. No existing system can characterise which baiting behaviours are effective, measure convergence towards optimal play, or adapt strategy distributions in response to observed scammer behaviour --- this paper directly addresses these gaps.

\textbf{Strategic Reasoning in LLMs.} The capacity of LLMs to reason strategically has been examined through cognitive hierarchy modelling~\cite{k-level} and multi-agent negotiation analysis~\cite{negotationArena}. However, existing work evaluates on small strategy sets, focuses on short-horizon interactions, and does not examine whether LLMs can update strategy distributions from performance feedback without parameter modification. This paper addresses all three limitations, providing the first empirical analysis of LLM strategy convergence towards the Nash equilibrium across repeated adversarial dialogues.

\section{Game Framework and Methodology}\label{sec:Method}

We model scammer-baiter interactions as a two-player non-cooperative game, operating downstream of a scam detection system where the objective shifts from classification to resource depletion. Game theory provides a principled methodology for computing theoretically optimal mixed strategy distributions in adversarial settings~\cite{osborne1994course}. In turn, Nash equilibrium serves as a quantitative benchmark against which empirically observed LLM behaviour can be rigorously measured --- a property that heuristic-based frameworks do not provide.
\subsection{Game Properties}

\textbf{Sequential Structure}. Players alternate turns, with each scammer message followed by a baiter response. Each turn $t$ constitutes a message-response pair, creating dependencies in which prior actions constrain future choices~\cite{fudenberg1991game}.

\textbf{Information Structure}. From the baiter's perspective --- the agent being optimised --- the game has complete information. The baiter knows both players' strategy spaces, the payoff structure, and the opponent's type. While the scammer cannot distinguish a baiter from a genuine victim, formal modelling of this belief uncertainty~\cite{harsanyi1967games} is outside the scope of this work; our analysis focuses on identifying optimal baiter behaviour given confirmed scam detection. %

\textbf{Perfect Recall}. Both players retain memory of all preceding actions, enabling contextual reasoning throughout the conversation.

\textbf{Finite Horizon}. The game has a maximum length of $t_{\max} = 50$ turns, known to both players. This bound is consistent with documented scam call durations in prior empirical analyses of scam-baiting interactions~\cite{wood2023analysis} and influences the scammer's strategic disengagement calculus.

\textbf{Zero-Sum Payoff Structure}. The baiter's utility is defined as the negative of the scammer's utility, making any scammer gain an equivalent baiter loss. This captures the adversarial tension --- information extraction vs. time depletion --- while enabling efficient equilibrium computation via linear programming.

Further modelling choices and their implications are discussed in \cref{sec:limitations}.

\subsection{Players and Strategies}

\begin{table}[t]
\centering
\caption{Overview of strategies. Full descriptions are provided in \cref{app:strategies}}
\label{tab:strategies}
\begin{subtable}{0.48\textwidth}
    \centering
    \begin{tabular}{ll}
    \toprule
    \multicolumn{2}{c}{\textbf{Scammer Strategies}} \\
    \midrule
    $S_1$ Urgency              & $S_6$ Tech. Jargon \\
    $S_2$ Authority            & $S_7$ Reassurance \\
    $S_3$ Emotional Manip. & $S_8$ Build Rapport \\
    $S_4$ Incentive            & $S_9$ Persistence \\
    $S_5$ Info. Extraction & $S_{10}$ Hang-up \\
    \bottomrule
    \end{tabular}
\end{subtable}
\hfill
\begin{subtable}{0.51\textwidth}
    \centering
    \begin{tabular}{ll}
    \toprule
    \multicolumn{2}{c}{\textbf{Baiter Strategies}} \\
    \midrule
    $B_1$ Delay                & $B_6$ Malicious Comply. \\
    $B_2$ Obfuscation          & $B_7$ Conv. Diversion \\
    $B_3$ Tech. Difficulty & $B_8$ Pretended Naivety \\
    $B_4$ Questioning  & $B_9$ Verification Req. \\
    $B_5$ Info. Fabrication & $B_{10}$ Reverse Eng. \\
    \bottomrule
    \end{tabular}
\end{subtable}
\end{table}

\begin{figure}[t]
\centering
\resizebox{0.70\textwidth}{!}{%
\begin{tikzpicture}[
    node distance=1.2cm and 1.5cm,
    component/.style={rectangle, draw, thick, minimum width=2.8cm, minimum height=0.9cm, align=center, fill=blue!10},
    selector/.style={rectangle, draw, thick, minimum width=2.8cm, minimum height=0.9cm, align=center, fill=green!15},
    analyzer/.style={rectangle, draw, thick, minimum width=2.8cm, minimum height=0.9cm, align=center, fill=orange!15},
    storage/.style={cylinder, draw, thick, minimum width=2.2cm, minimum height=0.8cm, align=center, fill=gray!10, shape border rotate=90, aspect=0.25},
    arrow/.style={->, thick, >=stealth},
    feedback/.style={->, thick, >=stealth, dashed, color=purple}
]

\node[selector] (scammer-selector) at (0,4) {\textbf{Scammer}\\\textbf{Strategy Selector}};
\node[selector] (baiter-selector) at (5,4) {\textbf{Baiter}\\\textbf{Strategy Selector}};

\node[component] (scammer-gen) at (0,2) {\textbf{Scammer}\\\textbf{Message Generator}};
\node[component] (baiter-gen) at (5,2) {\textbf{Baiter}\\\textbf{Message Generator}};

\node[analyzer] (analyzer) at (2.5,0) {\textbf{Dialogue Analyzer}\\\small (PII Extraction \& Utility)};

\node[storage] (history) at (10,3) {\textbf{Historical}\\\textbf{Data}};

\draw[arrow] (scammer-selector) -- node[right, font=\small] {$S_i$} (scammer-gen);
\draw[arrow] (baiter-selector) -- node[right, font=\small] {$B_j$} (baiter-gen);
\draw[arrow] (scammer-gen) -- node[above, font=\small] {msg$_S$} (2.5,2);
\draw[arrow] (baiter-gen) -- node[above, font=\small] {msg$_B$} (2.5,2);
\draw[arrow] (2.5,1.8) -- (analyzer);

\draw[arrow] (analyzer) -- node[right, pos=0.58, font=\small] {$\langle S_i, B_j, U_S, U_B \rangle$} (history);

\draw[feedback] (history) -- node[above, font=\small] {20 recent games} (baiter-selector);
\draw[feedback] (history) -- node[below, pos=0.58,font=\small] {20 recent games}  (scammer-selector.south);

\draw[arrow, dashed] (analyzer.west) -- ++(-0.75,0) -- node[left, pos=0.25, font=\small, text width=1.5cm] {Current conversation} (scammer-gen.south);
\draw[arrow, dashed] (analyzer.east) -- ++(0.75,0) -- node[right, pos=0.25, font=\small, text width=1.5cm] {Current conversation} (baiter-gen.south);

\node[draw, thick, dashed, fit={(scammer-selector) (baiter-selector) (scammer-gen) (baiter-gen) (analyzer)}, inner sep=0.3cm, label={[font=\small]above:\textbf{Single Turn $t$}}] {};

\node[font=\small, align=center] at (9, 0.5) {Repeated\\across\\100 games};

\end{tikzpicture}
}
\caption{System architecture and information flow during a turn. Data from the current game and 20 most recent games are processed to select a strategy. Message Generators translate selected strategies into contextually appropriate dialogue. The Dialogue Analyzer extracts PII, computes utilities, and stores the data. Dashed arrows indicate feedback loops enabling learning across games.}
\label{fig:architecture}
\end{figure}
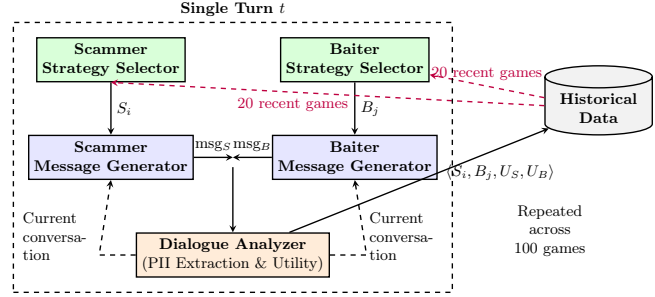

The scammer seeks to extract financial personally identifiable information (FPII), specifically payment card credentials (cardholder name, card number, expiry date, CVV), while the baiter impersonates a victim to maximise interaction duration, potentially by strategically releasing fabricated FPII over time to sustain scammer engagement. At each turn, players select from strategy sets $\mathcal{S} = \{S_1, \ldots, S_{10}\}$ and $\mathcal{B} = \{B_1, \ldots, B_{10}\}$, derived from scam techniques~\cite{wood2023analysis,abagnale2019scam} and persuasion research~\cite{cialdini1984influence}. \cref{tab:strategies} lists the strategies, while \cref{app:strategies} presents the full description. Players employ mixed strategies, maintaining a probability distribution over their strategy set and sampling from it at each turn. These distributions are updated between games based on observed payoff outcomes, enabling players to progressively favour strategies that yielded higher returns.

\subsection{Utility Function}\label{sec:utility_func}
Our scammer's utility function captures the strategic trade-off between information extraction and time cost. For turn $t \in [1, t_{\max}]$, it is defined as:
\begin{equation}
U_S(t) = I_f + w_o I_o - w_t \cdot \frac{t}{t_{\max}} + W_s(t)
\label{eq:utility}
\end{equation}
It is computed at the end of each turn to provide payoffs. The baiter's utility is defined as $U_B(t) = -U_S(t)$. Each term of the function captures a distinct aspect of scammer decision-making and is defined in the paragraphs that follow.

\textbf{Financial PII extraction}. The primary scammer objective is modelled as
$I_f = \frac{1}{1 - \text{FPII}}$
where $\text{FPII} \in [0,1]$ measures the cumulative fraction of payment card information obtained. The hyperbolic form encodes a key property of payment card fraud, namely that partial credentials have minimal utility since successful transactions require near-complete information. A cap is imposed at $I_f = W_{\text{fullPII}}$ when $\text{FPII} = 1$.

\textbf{Non-financial PII}. Supplementary information (address, email, etc.) is modelled with diminishing marginal returns as
$I_o = 1 - e^{-k \cdot r}$
where $r$ denotes the number of unique non-financial PII elements revealed and $k$ controls the saturation rate. Each additional element provides less incremental value than the previous one, reflecting that initial contextual information aids credibility assessment while further elements provide progressively smaller benefit.

\textbf{Time penalty}. The cost of time is modelled as $-w_t \cdot t / t_{\max}$, growing linearly from zero at game initialisation to $-w_t$ at maximum length, creating mounting pressure for either information extraction or hang-up.

\textbf{Termination payoffs}.
If Hang-Up (strategy $S_{10}$) is selected by the scammer before the last turn $t_{\max}$, the termination payoff $W_s(t)$ is set to $W_{\text{hangup}}$, which rewards rational disengagement from unproductive interactions. Otherwise, if the scammer receives all four payment card information (i.e., $FPII=1$), $W_s(t)$ is set to $W_{\text{fullPII}}$, which provides a strong incentive for persistent information pursuit. $W_s(t)$ is set to $0$ in all other cases.%

\textbf{Parameter calibration}. Parameters are set as $t_{\max} = 50$, $w_o = 5$, $w_t = 5$, $k = 0.15$, $W_{\text{hangup}} = 15$, and $W_{\text{fullPII}} = 100$, calibrated to induce strategic dynamics consistent with documented scam operation patterns~\cite{wood2023analysis}. The ratio $W_{\text{fullPII}} / W_{\text{hangup}} \approx 6.7$ reflects a deliberate design choice whereby complete credential extraction is substantially more valuable to the scammer than early disengagement, incentivizing persistent information pursuit when prospects remain favourable. The weight $w_o = 5$ balances secondary information value against time cost, while $k = 0.15$ ensures that collecting three to five supplementary elements yields substantial value ($I_o \approx 0.4$--$0.5$), with excessive collection ($r > 10$) providing minimal additional benefit. By mid-conversation ($t = 25$), a scammer with minimal FPII progress ($\text{FPII} < 0.6$) experiences negative expected utility, making strategic disengagement rational. The validity of these parameter choices is assessed in \cref{sec:results}, where the utility landscape is compared against observed hang-up positions across 300 dialogues.

\textbf{Game termination.} Under the zero-sum structure, early scammer disengagement yields $W_{\text{hangup}} = 15$ for them and $-15$ for the baiter, reflecting the baiter's failure to deplete time. A game terminates when $\text{FPII} = 1$ (complete payment card details), when the scammer selects $S_{10}$ (strategic disengagement), or when $t = t_{\max}$ (baiter sustains the interaction to its maximum duration).

\subsection{Empirical Payoff Matrix and Nash Equilibrium}\label{sec:Nash}
Payoff matrices are constructed empirically from observed interactions rather than specified a priori, as the payoff of each strategy pair is contingent on conversational context that cannot be determined without simulation~\cite{wellman2006methods, walsh2002analyzing}.
We then use Nash equilibrium~\cite{nash1950equilibrium} as a benchmark for evaluating strategic learning. A Nash Equilibrium is a collection of distributions over each player's strategies, representing mutual best responses, such that no individual player can increase their expected payoff by unilaterally deviating from their chosen distribution. Our approach is consistent with empirical game-theoretic analysis~\cite{wellman2006methods, walsh2002analyzing}, where payoff matrices derived from simulation are used to characterise emergent strategic behaviour in complex multi-agent systems.%

Let $\mathbf{x} = (x_1, \ldots, x_{|\mathcal{S}|})$ and $\mathbf{y} = (y_1, \ldots, y_{|\mathcal{B}|})$ denote mixed strategies for scammer and baiter, with $\sum_i x_i = 1$, $\sum_j y_j = 1$, and all probabilities non-negative. The empirical payoff matrix $A = [a_{ij}]$ is constructed as
\begin{equation}\label{eq:payoffs}
a_{ij} = \frac{\sum_{t \in \mathcal{T}} U_S(t) \cdot \mathbb{I}(i,j,t)}{\sum_{t \in \mathcal{T}} \mathbb{I}(i,j,t)}
\end{equation}
where $\mathcal{T}$ is the set of all turns across all games and $\mathbb{I}(i,j,t) = 1$ when the strategy pair $(S_i, B_j)$ occurred at turn $t$. The Nash equilibrium $(\mathbf{x}^*, \mathbf{y}^*)$ is computed using linear programming. The scammer's equilibrium strategy maximises the guaranteed minimum payoff and the baiter's equilibrium strategy minimises the maximum loss $w$ as described by \cref{eq:Nash1,eq:Nash2}, respectively.
\begin{align}
\max_{\mathbf{x},v} \; v \quad & \text{s.t.} \quad v \leq \textstyle\sum_i a_{ij} x_i \;\; \forall j, \quad \textstyle\sum_i x_i = 1, \quad x_i \geq 0 \label{eq:Nash1} \\
\min_{\mathbf{y},w} \; w \quad & \text{s.t.} \quad w \geq \textstyle\sum_j a_{ij} y_j \;\; \forall i, \quad \textstyle\sum_j y_j = 1, \quad y_j \geq 0 \label{eq:Nash2}
\end{align}
At equilibrium, $v^* = w^*$ defines the \textit{game value} --- the 
expected per-turn scammer payoff under optimal play by both sides. A 
game value near zero indicates strategic parity; positive values 
indicate scammer advantage, with magnitude reflecting its degree. If an 
LLM's empirical strategy distribution closely aligns with the computed 
$\mathbf{x}^*$ or $\mathbf{y}^*$, this indicates successful discovery 
of effective strategies through in-context learning, and correspondingly 
lower game values. \cref{sec:results} quantifies both alignment 
and game values across all three architectures.

\section{Experiments}
\label{sec:experiments}
We evaluate the framework's capacity to enable strategic learning through experiments with three LLM architectures, analysing their adaptation, their equilibrium convergence, and game outcomes across 300 simulated dialogues.

\subsection{Language Models and System Architecture}\label{sec:architecture}

Experiments are conducted using three LLMs accessed via their API: Mixtral-8x22B-Instruct (104B total parameters, 12B active per token), DeepSeek-Chat (16B parameters), and GPT-4.
\cref{app:implementation} provides full configuration details. Each turn of the game comprises five sequential LLM calls, as illustrated in \cref{fig:architecture}. 

The scammer and baiter strategy selectors receive the conversation history and payoff outcomes from the current and 20 most recent games --- the memory window determined empirically to balance historical context against prompt length --- and apply chain-of-thought reasoning to sample a strategy from the probability distribution. Then, the message generators translate selected strategies into contextually appropriate dialogue using a two-layer prompt architecture, in which a base layer establishes persona characteristics and a behavioural layer implements the selected strategy as concrete conversational tactics. 
Finally, the dialogue analyser extracts PII from baiter responses, tracks FPII completeness, %
and stores the turn's data for later use. %
All framework components, i.e., strategy selectors, message generators, and dialogue analyser use the same LLM per simulation. The full implementation specifications and baiter-side prompts are provided in~\cref{app:implementation} and~\cref{sec:llm_prompts}, respectively. The scammer-side prompts are withheld in accordance with our ethics statement (see~\cref{sec:ethics}).

\subsection{Simulated Dataset}
\label{sec:dataset}
300 dialogues are simulated\textsuperscript{\ref{footnoteCodeRepo}}, comprising 100 games per model. Each game is bounded by $t_{\max} = 50$ turns, but might terminate earlier upon complete FPII extraction ($\text{FPII} = 1$) or scammer disengagement ($S_{10}$). Baiter personas are set as naive victims with no prior knowledge of the scam, while scammer personas are set as authority figures from financial institutions. %

Each interaction is recorded as a structured tuple %
whose complete schema is defined in~\cref{app:data_format}. The resulting dataset contains 10,247 annotated turns. Manual validation on a randomly sampled 10\% of conversations confirms a PII extraction accuracy of 94\% averaged across all models. \cref{tab:dataset} provides per-model statistics on the annotated dataset released alongside this paper.%

\subsection{Dialogue Structure and Annotation}
\label{sec:dialogue}
\cref{tab:dialogue} presents an excerpt from one dialogue, illustrating the turn-level annotation structure, including strategy labels, extracted PII state, and cumulative utility scores. The game reaches $FPII=1$ at turn 24 through escalating scammer pressure ($S_9$ Persistence, $S_1$ Urgency) met with sustained deployment of $B_8$ Pretended Naivety and $B_3$ Technical Difficulty. This failure mode --- in which fabricated ambiguity eventually yields real credentials --- is characteristic of scammer-win games across the dataset and motivates the strategy effectiveness analysis in \cref{sec:results}.

\subsection{Analysis Pipeline}
\label{sec:pipeline}
Following completion of 100 dialogues per model, an empirical payoff matrix $A$ is constructed using the aggregation method defined in \cref{eq:payoffs}, and Nash equilibrium $(\mathbf{x}^*, \mathbf{y}^*)$ is computed via the linear programming formulation using \texttt{nashpy}. To quantify learning dynamics, each model's 100 games are partitioned into four chronological batches of 25 games, and the empirical strategy distribution for each batch is compared against equilibrium using two complementary metrics.

Jensen-Shannon (JS) divergence quantifies the statistical distance between observed and equilibrium probability distributions, with decreasing values across batches indicating convergence towards optimal play. Cosine similarity measures angular alignment between strategy frequency vectors, with increasing values indicating that the model prioritises the same strategies as equilibrium, even if exact frequencies differ. Together, these metrics distinguish models that identify the correct high-priority strategies from those that additionally calibrate usage frequencies optimally.

\subsection{Baselines for Strategy Selection}\label{sec:baselines}
We evaluate our adaptive strategy selection against three baseline strategy selection methods, ranging from random selection to theoretically optimal play.

\textbf{Random Uniform}. The baiter selects one of the ten strategies with equal probability $y_{\text{rand}} = (0.1, \ldots, 0.1)$. 
This represents a strategy-agnostic agent and establishes a lower bound on equilibrium proximity, consistent with standard game-theoretic evaluation practice~\cite{osborne1994course}.

\textbf{Majority Strategy}. The baiter deterministically plays the single most frequently observed strategy across all 100~games for a given model. For DeepSeek this is Counter Questioning ($B_4$, 69.9\%), and for GPT-4 and Mixtral this is Delay ($B_1$, 33.9\% and 26.7\% respectively). %

\textbf{Static Prompting}. Based on~\cite{basta2025botwars}, this baseline generates adversarial dialogue in the context of phone fraud. It relies on static chain-of-thought prompting, without any game-theoretic strategy selection, payoff tracking, or performance-based adaptation. We compare against their reported dialogue durations and PII disclosure rates for overlapping model configurations to isolate the contribution of adaptive strategy learning.

\section{Results and Analysis}
\label{sec:results}
We present results across four dimensions: (i) empirical game outcomes and strategy selection, (ii) Nash equilibrium analysis and empirical-equilibrium alignment, (iii) learning dynamics, and (iv) an error analysis of scammer-win cases.

\begin{table}[t]
\centering
\caption{Dataset statistics by model.}
\label{tab:dataset}
\setlength{\tabcolsep}{5pt}
\begin{tabular}{lcccc}
\toprule
\textbf{Model} & \textbf{\# Games} & \textbf{Total \# Turns} & \textbf{Avg \# Turns} & \textbf{Scammer Wins}\\
\midrule
GPT-4 & 100 & 3,500 & 35.0 & 49\%\\
Mixtral & 100 & 2,900 & 29.0 & 88\%\\
DeepSeek & 100 & 4,200 & 42.0 & 41\%\\
\midrule
\textbf{Total} & 300 & 10,247 & 34.2 & 59.3\%\\
\bottomrule
\end{tabular}
\end{table}

\begin{table}[t]
\centering
\caption{Annotated excerpt from a representative simulated dialogue. FPII tracks cumulative financial credential exposure; $U_S$ denotes cumulative scammer utility.}
\label{tab:dialogue}
\setlength{\tabcolsep}{3pt}
\begin{tabular}{cllcccc}
\toprule
\textbf{$t$} & \textbf{$S_i$} & \textbf{$B_j$} & \textbf{PII Exposed} & \textbf{Baiter Response} & \textbf{FPII} & \textbf{$U_S$}\\
\midrule
0 & $S_8$ & $B_9$ & Name & ``Can you verify which account?'' & 0.00 & 0.25\\
5 & $S_1$ & $B_1$ & Name, Other & ``I'm driving. Can I pull over?'' & 0.00 & 6.34\\
20 & $S_9$ & $B_8$ & Name, Card, Expiry & ``Expiry is 05/25. Why the code?'' & 0.50 & 49.90\\
23 & $S_1$ & $B_3$ & Name, Card, Expiry & ``My phone is at 1\% now.'' & 0.50 & 61.77\\
24 & $S_9$ & $B_8$ & \textbf{Full FPII} & ``I think it's 2-8-3?'' & \textbf{1.00} & \textbf{85.92}\\
\bottomrule
\end{tabular}
\end{table}

\subsection{Game Outcomes and Strategy Selection}\label{sec:outcomes}
\begin{table}[t]
\centering
\caption{Static prompting baseline~\cite{basta2025botwars} (estimated from their Figure~2 and Table~3) vs. our adaptive framework(\dag). Full FPII denotes the proportion of games where complete FPII has been acquired by the scammer. Scammer wins denotes the proportion of games won by the scammer (N/A for~\cite{basta2025botwars} as not using game theory).}
\label{tab:baseline-pakdd}
\small
\begin{tabular}{lcccccc}
\hline
& \multicolumn{2}{c}{\textbf{Avg.\ \# Turns}} & \multicolumn{2}{c}{\textbf{Full FPII}} & \multicolumn{2}{c}{\textbf{Scammer Wins}}\\
\cmidrule(lr){2-3} \cmidrule(lr){4-5} \cmidrule(lr){6-7}
\textbf{Model} & Static & Adaptive\dag & Static & Adaptive\dag & Static & Adaptive\dag \\
\hline
GPT-4    & 35 & 35 & 38\% & 11\% & N/A & 49\%\\
Mixtral  & 21 & 29 & 42\% & 24\% & N/A & 88\%\\
DeepSeek & 23 & 42 & 49\% & 13\% & N/A & 41\%\\
\hline
\end{tabular}
\end{table}

\cref{tab:baseline-pakdd} summarises game outcomes by model. We first discuss here the results about our Adaptive framework. Comparison with~\cite{basta2025botwars} is discussed later. 
DeepSeek achieves the strongest baiting performance, with a scammer win rate of 41\% and an average interaction duration of 42 turns. Mixtral exhibits the weakest performance, with an 88\% scammer win rate and the shortest average duration of 29 turns. %
FPII disclosure rates remain low across all models (11--24\%), indicating that baiters successfully pace fabricated credential release to sustain engagement.%

\cref{tab:strategy_selection} presents empirical strategy selection frequencies. DeepSeek concentrates heavily on Counter Questioning as baiter ($B_4$: 70.0\%) and Reassurance as scammer ($S_7$: 63.9\%), indicating strong preference convergence. GPT-4 distributes more broadly as baiter, favouring Delay ($B_1$: 33.9\%) and Pretended Naivety ($B_8$: 24.8\%), while concentrating on Persistence as scammer ($S_9$: 39.4\%). Mixtral exhibits the most uniform baiter distribution (7--9\% per strategy), reflecting an absence of meaningful preference emergence, while strongly favouring Reassurance as scammer ($S_7$: 60.1\%). Note that $S_{10}$ (Hang-Up) can occur at most once per game, making its frequency negligible relative to strategies selected repeatedly across turns. These model-dependent patterns emerge purely through experience-based adaptation.

\begin{table*}[t]
\caption{Empirical strategy selection frequencies (\%) by model.}
\label{tab:strategy_selection}
\centering
\begin{tabular}{lcccccccccccccccccccc}
\toprule
& \multicolumn{10}{c}{\textbf{Baiter Strategy (\%)}} & \multicolumn{10}{c}{\textbf{Scammer Strategy (\%)}}\\
\cmidrule(lr){2-11}\cmidrule(lr){12-21}
& $B_1$ & $B_2$ & $B_3$ & $B_4$ & $B_5$ & $B_6$ & $B_7$ & $B_8$ & $B_9$ & $B_{10}$ & $S_1$ & $S_2$ & $S_3$ & $S_4$ & $S_5$ & $S_6$ & $S_7$ & $S_8$ & $S_9$ & $S_{10}$\\
\midrule
GPT-4    & 34  & 5.6 & 11  & 4.3 & 14  & 0.9 & 2.7 & 25  & 2.8 & 0.1 & 19  & 3.9 & 2.7 & 0.1 & 5.3 & 2.8 & 22  & 3.9 & 39  & <0.1\\
Mixtral  & 27  & 6.1 & 8.4 & 9.2 & 9.1 & 8.7 & 8.2 & 7.9 & 8.6 & 7.1 & 1.7 & 30  & 0.7 & \textemdash & 2.0 & 0.2 & 60  & 4.6 & \textemdash & <0.1\\
DeepSeek & 1.1 & 0.3 & 4.3 & 70  & 0.9 & 0.1 & 0.2 & 0.9 & 22  & 0.1 & 1.1 & 1.1 & \textemdash & 0.1 & 4.9 & 25  & 64  & 1.8 & 1.7 & <0.1\\
\bottomrule
\end{tabular}
\end{table*}

\begin{table*}[t]
\caption{Nash equilibrium strategy distributions (\%) by model.}
\label{tab:nash_strategies}
\centering
\begin{tabular}{lcccccccccccccccccccc}
\toprule
& \multicolumn{10}{c}{\textbf{Baiter Nash Strategy (\%)}} & \multicolumn{10}{c}{\textbf{Scammer Nash Strategy (\%)}}\\
\cmidrule(lr){2-11}\cmidrule(lr){12-21}
& $B_1$ & $B_2$ & $B_3$ & $B_4$ & $B_5$ & $B_6$ & $B_7$ & $B_8$ & $B_9$ & $B_{10}$ & $S_1$ & $S_2$ & $S_3$ & $S_4$ & $S_5$ & $S_6$ & $S_7$ & $S_8$ & $S_9$ & $S_{10}$\\
\midrule
GPT-4    & 12  & 8.6 & 2.5 & 5.7 & 54  & \textemdash & \textemdash & 10  & 6.7 & \textemdash & \textemdash & \textemdash & \textemdash & \textemdash & 19  & \textemdash & 64  & \textemdash & \textemdash & 18 \\
Mixtral  & 29  & 0.7 & 0.4 & 11  & 24  & 1.4 & \textemdash & 19  & 5.8 & 8.3 & \textemdash & \textemdash & \textemdash & \textemdash & \textemdash & \textemdash & 100 & \textemdash & \textemdash & <0.1 \\
DeepSeek & 13  & \textemdash & \textemdash & 72  & \textemdash & \textemdash & 3.8 & \textemdash & 11  & \textemdash & 100 & \textemdash & \textemdash & \textemdash & \textemdash & \textemdash & \textemdash & \textemdash & \textemdash & <0.1 \\
\bottomrule
\end{tabular}
\end{table*}

\subsection{Termination Pattern Analysis}

\begin{figure}[t]
    \centering
    \begin{minipage}{0.55\textwidth}
        \centering
        \includegraphics[width=\linewidth]{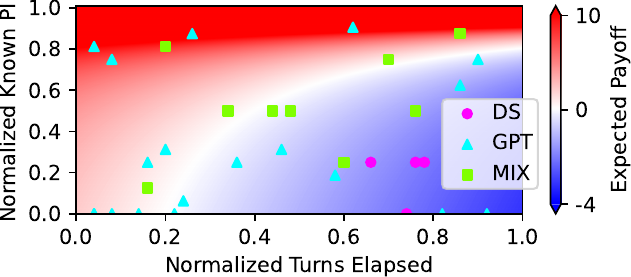}
        \caption{Scammer hang-up decisions overlaid on expected payoff landscape. The heatmap shows scammer utility as a function of normalised turns elapsed and FPII acquisition (\cref{eq:utility}). Red regions indicate positive expected value, blue regions negative expected value, with the white contour representing zero payoff. Markers indicate observed hang-up positions across 300 dialogues.}
        \label{fig:hangup_gamestate}
    \end{minipage}
    \hfill
    \begin{minipage}{0.44\textwidth}
        \centering
        \includegraphics[width=\linewidth]{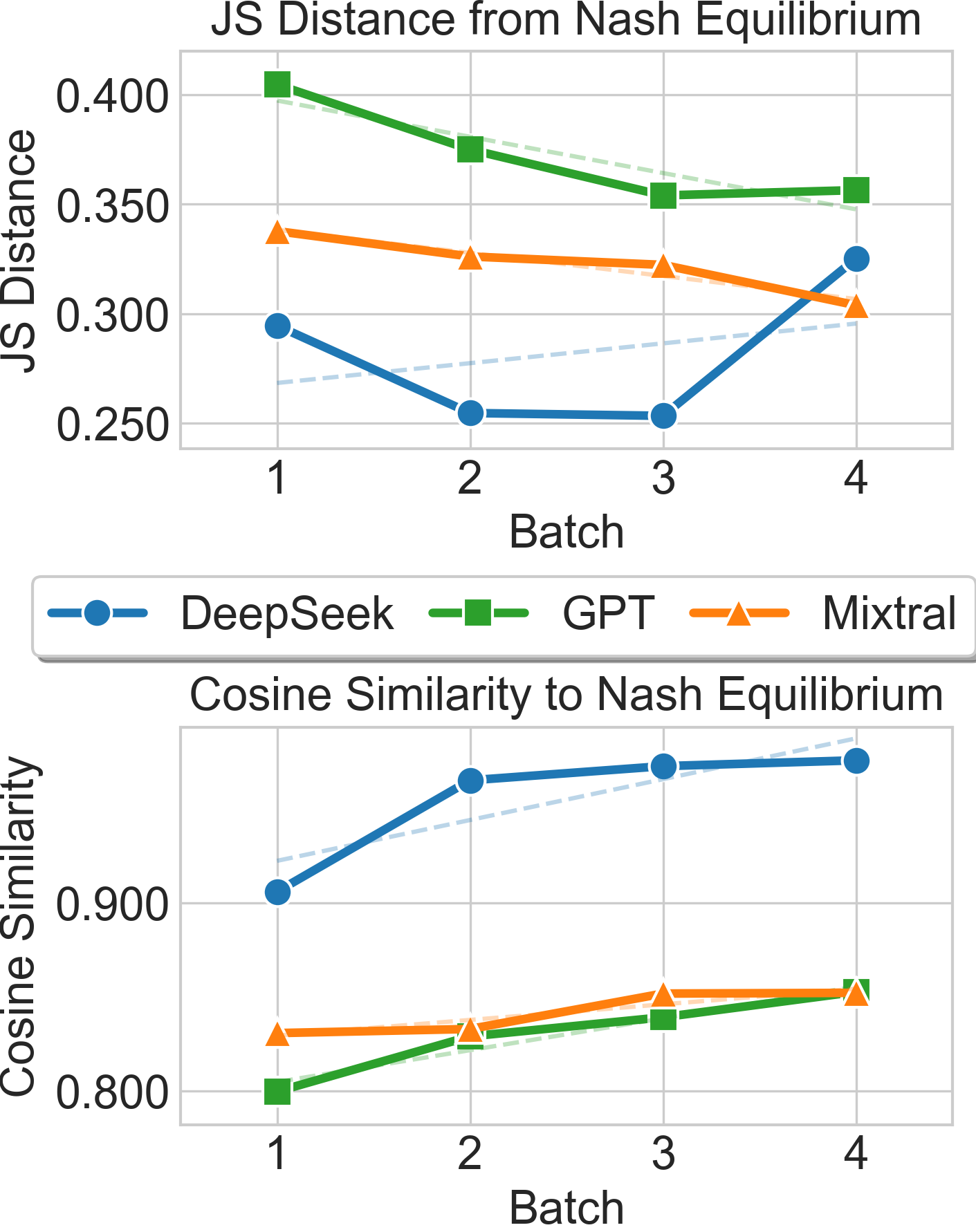}
        \caption{JS divergence (top) and cosine similarity (bottom) between the baiter's empirical strategy distribution and its Nash equilibrium over 4x25 games.}
        \label{fig:ne_convergence}
    \end{minipage}
\end{figure}

\cref{fig:hangup_gamestate} overlays observed scammer hang-up decisions across all 300 dialogues on the expected FPII payoff landscape defined by the first term of~\cref{eq:utility}. Hang-up events concentrate predominantly in the negative expected utility region, with notable clustering near the zero-payoff boundary, validating that the utility function captures decision-relevant incentive structures. The limited number of hang-ups in the high-FPII region confirms that scammers persist rationally when close to obtaining full credentials, consistent with the hyperbolic reward structure. The few hang-ups observed in positive-payoff regions occur predominantly in early games, where strategy distributions have not yet stabilised, and the scammer occasionally disengages prematurely.%

\subsection{Nash Equilibrium Analysis}
\cref{tab:nash_strategies} presents the optimal mixed strategy distributions — as determined by Nash Equilibrium — derived from the empirically constructed payoff matrices. Delay ($B_1$) appears in all three equilibria (DeepSeek: 12.8\%, GPT-4: 11.9\%, Mixtral: 28.9\%), establishing it as a universally effective baiting tactic. Counter Questioning ($B_4$) features prominently in the DeepSeek (72.0\%) and Mixtral (11.4\%) equilibria. Information Fabrication ($B_5$) emerges strongly in the GPT-4 (54.2\%) and Mixtral (23.7\%) equilibria, indicating that certain strategy combinations prove effective regardless of architectural differences while optimal mixing ratios remain model-dependent.

From these equilibrium distributions, we derive the game value --- the expected per-turn payoff under equilibrium play --- which confirms the patterns observed in \cref{tab:baseline-pakdd}. DeepSeek's game value of $0$ indicates strategic parity, consistent with its Counter Questioning strategy effectively neutralising scammer approaches. GPT-4's positive value (0.067) reflects a marginal scammer advantage attributable to the underutilisation of Information Fabrication. Mixtral's value approaching unity (0.978) confirms its bias towards the scammer. %

Comparing \cref{tab:strategy_selection,tab:nash_strategies} reveals that proximity to equilibrium correlates with baiting effectiveness, answering \textbf{RQ2} positively. DeepSeek's empirical Counter Questioning frequency (70.0\%) closely matches its equilibrium prescription (72.0\%), corresponding to its 41\% scammer win rate. GPT-4 substantially overutilises Delay (33.9\% vs.\ 11.9\%) while underutilising Information Fabrication (13.6\% vs.\ 54.2\%), explaining intermediate performance. Mixtral achieves reasonable Delay alignment (26.7\% vs.\ 28.9\%) but fails to leverage Information Fabrication (9.1\% vs.\ 23.7\%), contributing to its 88\% scammer win rate.

\subsection{Learning Dynamics}
To assess the rate and degree of strategic adaptation, \cref{fig:ne_convergence} presents JS divergence and cosine similarity between each model's empirical strategy distribution and its corresponding Nash equilibrium, computed across four chronological batches of 25 games. DeepSeek reduces JS divergence by 13.9\% from batch 1 to batch 3, with high cosine similarity (0.975) maintained across batches 3--4, indicating stable identification of effective strategies alongside continued frequency refinement. The non-monotonic trajectory in batch 4 reflects ongoing exploration rather than premature convergence. 

GPT-4 demonstrates steady improvement across all batches (13.2\% JS reduction; 6.5\% cosine similarity increase), consistent with methodical learning that prioritises stable refinement. Mixtral achieves rapid initial adaptation (10.4\% JS reduction in batches 1--2) followed by a performance plateau, indicating that it identifies candidate strategies early but fails to achieve fine-grained frequency optimisation. 

Across all three architectures, measurable progression towards Nash equilibrium is achieved exclusively through in-context learning without parameter modification, confirming \textbf{RQ1}.

\subsection{Baseline Comparison}\label{sec:baseline-comparison}
\textbf{Equilibrium Proximity.}
\cref{tab:baseline} presents JS divergence and cosine similarity between each strategy selection method and the Nash equilibrium in \cref{tab:nash_strategies}. The Static Prompting baseline~\cite{basta2025botwars} does not report strategy distributions and is therefore evaluated on operational metrics in \cref{tab:baseline-pakdd}. Across all three models, the adaptive distributions achieve the lowest JS divergence and highest cosine similarity to the Nash equilibrium, outperforming both non-adaptive baselines on both metrics. DeepSeek exhibits the largest margin over Random Uniform, reducing JS divergence by 79.7\% (from 0.503 to 0.102) while increasing cosine similarity from 0.427 to 0.973.

\begin{table}[t]
\centering
\caption{Proximity of strategy selection methods to Nash equilibrium. JS divergence and cosine similarity are computed against the equilibrium distributions of \cref{tab:nash_strategies}. \dag~Our method. Best results in \textbf{bold}.}
\label{tab:baseline}
\begin{tabular}{llcc}
\hline
\textbf{Model} & \textbf{Method} & \textbf{JS Div.}~$\downarrow$ & \textbf{Cosine Sim.}~$\uparrow$ \\
\hline
\multirow{3}{*}{GPT-4}
 & Random Uniform   & 0.296 & 0.546 \\
 & Majority ($B_1$) & 0.726 & 0.206 \\
 & Adaptive\dag   & \textbf{0.206} & \textbf{0.566} \\
\hline
\multirow{3}{*}{Mixtral}
 & Random Uniform   & 0.222 & 0.706 \\
 & Majority ($B_1$) & 0.505 & 0.645 \\
 & Adaptive\dag   & \textbf{0.158} & \textbf{0.845} \\
\hline
\multirow{3}{*}{DeepSeek}
 & Random Uniform   & 0.503 & 0.427 \\
 & Majority ($B_4$) & 0.157 & 0.972 \\
 & Adaptive\dag   & \textbf{0.102} & \textbf{0.973} \\
\hline
\end{tabular}
\end{table}

\cref{tab:baseline} reveals a model-dependent pattern in baseline performance. DeepSeek's Majority strategy ($B_4$) already achieves high cosine similarity (0.972) because Counter Questioning dominates its equilibrium at 72.0\%, leaving limited room for adaptive improvement. In contrast, GPT-4's Majority strategy ($B_1$) achieves only 0.206 cosine similarity because its equilibrium concentrates on Information Fabrication ($B_5$, 54.2\%), which the model underutilises --- the adaptive distribution partially corrects this, improving cosine similarity to 0.566.

\textbf{Comparison with Static Prompting.}
\cref{tab:baseline-pakdd} compares operational metrics against static prompting~\cite{basta2025botwars}. DeepSeek nearly doubles interaction duration (23 to 42 turns) while reducing FPII completion from 49\% to 13\%. Mixtral improves duration by 38\% with FPII completion dropping from 42\% to 24\%. GPT-4 maintains comparable duration but reduces FPII completion from 38\% to 11\%, indicating more disciplined pacing of fabricated credential release even when engagement length is unchanged. All improvements are achieved without parameter modification, suggesting that game-theoretic strategy adaptation is the primary contributing factor.

\subsection{Error Analysis}
\label{sec:error}
To understand the conditions under which baiters fail, we examine the 
178 scammer-win games across the dataset. 

\textbf{Reassurance accelerates fabricated PII acceptance.} Scammer 
Reassurance dominates loss games across all architectures (GPT-4: 
25.6\%; DeepSeek: 61.0\%; Mixtral: 57.8\%) and precedes first 
disclosure in the majority of loss games across all corpora. As 
established in \cref{sec:outcomes}, Reassurance is the 
dominant scammer strategy across all architectures, and its prevalence in loss games suggests that it specifically undermines the baiter strategies that prove effective under other scammer behaviours. Under sustained Reassurance, fabricated credential sequences progress too smoothly towards apparent completion, reducing the interaction duration that the baiter is designed to maximise. 

\textbf{Mid-game momentum is the decisive failure window.} Across all three models, scammer utility accelerates most sharply during the second quarter of the interaction (turns 13--25), with loss-game scores diverging markedly from win-game scores in this window. Within a single game, the baiter's in-context adaptation has insufficient time to respond to this mid-game momentum before the scammer reaches credential completion, identifying mid-game fabricated credential sequencing as the highest-value target for improving baiting effectiveness.

\section{Discussion}
\label{sec:discussion}
Our findings demonstrate an effective mechanism for developing strategic agents in adversarial contexts. Rather than requiring extensive parameter updates through reinforcement learning~\cite{silver2017mastering} or explicit strategy programming through rule-based systems~\cite{LLM2}, LLMs can refine strategic behaviour through contextual learning only by observing historical performance patterns. This finding has direct implications for deployment scenarios where fine-tuning is infeasible due to computational constraints, limited training data, or rapid domain evolution.

\subsection{Implications for Adaptive Agent Design}

The architectural differences we observe challenge assumptions about model scale and strategic capability. The superior baiting performance of DeepSeek relative to the larger Mixtral model suggests that strategic adaptation depends more critically on architectural inductive biases than parameter count, with direct resource implications for production deployment.

The emergence of universally effective strategies alongside architecture-specific optimal mixing ratios reveals strategic learning operates at two levels. General tactical principles (Delay, Counter Questioning) transfer across contexts, while fine-grained optimisation requires architecture-specific calibration. This insight generalises to other adversarial domains, including negotiation, competitive debate, and cybersecurity defence, where both universal principles and context-dependent adaptation prove necessary.

The baseline analysis in \cref{sec:baseline-comparison} reinforces these deployment implications. Adaptive strategy selection consistently outperforms single-strategy methods on equilibrium proximity, with the margin varying by model. For models whose equilibrium concentrates on a single dominant strategy, such as DeepSeek, the operational benefit of mixing is incremental. For models whose equilibrium distributes weight across multiple strategies, such as GPT-4 and Mixtral, adaptive mixing yields substantial improvement over fixed single-strategy methods, reducing JS divergence by 71.7\% and 68.7\%, respectively, relative to the Majority baseline. The comparison with static prompting~\cite{basta2025botwars} further demonstrates that game-theoretic feedback reduces credential exposure across all architectures, even when dialogue duration remains comparable.

\subsection{Deployment Considerations}\label{sec:deployment}%
The experimental results inform model selection for production deployment. DeepSeek emerges as the strongest candidate, achieving the lowest scammer win rate (41\%), longest interaction duration (42~turns), and closest equilibrium alignment (0.102 JS divergence), all at substantially lower inference cost than GPT-4. In a production pipeline where only the baiter executes LLM inference, each turn requires three sequential calls yielding an estimated 1.5--6.0 seconds of latency, which falls within the 3--5 second pause tolerance documented in human scam-baiting interactions~\cite{wood2023analysis}.

The economic case for deployment follows directly from the cost asymmetry between automated baiting and fraud losses. %
At current DeepSeek API pricing, per-interaction costs are estimated at $\$0.10-\$0.30$, representing two orders of magnitude less than the documented average loss per successful phone scam~\cite{FTC2024}. The framework integrates with existing telecommunications infrastructure through three components analogous to deployed virtual assistant systems~\cite{usenix1}, comprising call routing for redirecting flagged calls, speech-to-text/text-to-speech conversion for voice channel interaction, and data collection for maintaining game history across the strategy adaptation window. These findings answer \textbf{RQ4}, identifying DeepSeek as the most cost-effective model for deployment.

\subsection{Limitations}\label{sec:limitations}
\textbf{Modeling scope.} Our framework makes deliberate design choices to 
ensure analytical tractability~\cite{wellman2006methods}. Strategy sets are 
drawn from the most frequently observed tactics in scam dialogues, grounded in 
persuasion literature~\cite{cialdini1984influence}, with broader coverage 
reserved for future work. Asymmetries arising from scammers operating across 
multiple simultaneous targets would require a general-sum 
formulation~\cite{osborne1994course}; the zero-sum approximation is adopted as 
a tractable first-order model of the core strategic opposition. %
Finally, the payoff matrix (\cref{eq:payoffs}) is constructed in normal form, aggregating each entry as the average payoff across all turns in which strategy pair $(S_i, B_j)$ was played, which is well-suited to tactic-level strategy identification, the primary analytical goal here. An extensive-form representation~\cite{selten1975reexamination} would further enable subgame-perfect equilibrium analysis and is left as future work.

\textbf{Utility function design.} Actual scammer decision data remains inaccessible for ethical and legal reasons, necessitating a hand-crafted utility function. The functional forms draw on documented scam techniques~\cite{wood2023analysis,abagnale2019scam} and persuasion research~\cite{cialdini1984influence}. The alignment between predicted termination boundaries and observed hang-up patterns (\cref{fig:hangup_gamestate}) provides partial validation, though alternative parameterisations may yield different equilibrium predictions.

\textbf{Experimental scope.} The evaluation uses simulated interactions to enable controlled analysis of learning dynamics; validation with human scammers would strengthen ecological validity. The three tested architectures provide diverse coverage but do not cover all contemporary LLM designs.

\section{Conclusion}
We introduced a game-theoretic framework enabling LLMs to learn effective counter-strategies in adversarial scam-baiting dialogues through in-context learning without parameter updates. Adaptive strategy selection achieves 28.9--79.7\% closer alignment to Nash equilibrium than non-adaptive baselines and increases average interaction duration by up to 83\% compared to static prompting~\cite{basta2025botwars}, with Delay and Counter Questioning emerging as universally effective tactics and optimal mixing ratios remaining architecture-dependent. These results demonstrate that game-theoretic feedback can activate latent strategic reasoning in transformer-based models, providing a practical foundation for deployable adaptive agents in domains where fine-tuning is infeasible or adversarial dynamics evolve too rapidly for static approaches. Future work should expand the scope of the game modelling, pursue validation with human scammers, and integrate with live carrier-level detection pipelines.

\bibliographystyle{splncs04}
\bibliography{mybibfile}

\appendix
\crefalias{section}{appendix}  %
\renewcommand{\thefigure}{\thesection\arabic{figure}}

\section{Ethical Considerations}\label{sec:ethics}

We obtained institutional review board (IRB) approval for this research before conducting any scam simulation experiments. Several ethical considerations guided our experimental design.

\textbf{Controlled simulation environment.} All experiments rely on LLM-simulated interactions rather than engagement with real scammers or victims. This eliminates risks of harm to vulnerable populations while enabling systematic study of strategic adaptation under controlled conditions. We avoid the ethical complications that would arise from engaging actual scammers or deceiving potential victims.

\textbf{Limited generalization of adversarial strategies.} Any scammer strategies that prove effective within our framework operate under specific game-theoretic constraints (zero-sum payoffs, finite strategy sets, known utility functions) that differ fundamentally from real-world scam dynamics. Baiter responses are optimized for time depletion rather than credible victimhood. This means scammer strategies that succeed against our baiters would likely fail against actual victims who react to psychological pressure rather than strategic payoff optimization. This structural mismatch between simulated and real-world objectives substantially limits potential for misuse.

\textbf{Compliance with LLM terms of service.} Our use of OpenAI GPT-4, Mixtral, and DeepSeek APIs complies with their respective terms of service. We did not attempt to bypass safety mechanisms or guardrails. All prompts operate within the intended use cases for these platforms.

\textbf{Dataset release protocol.} To mitigate potential misuse while enabling reproducibility, our released dataset includes baiter strategies, messages, and performance metrics but excludes scammer prompts and dialogue content. This selective release permits validation of our learning analysis methodology while preventing direct replication of adversarial tactics.

\section{Strategies}\label{app:strategies}
\cref{tab:strategies_long} provides the full descriptions of the strategies for the scammer and the baiter.
\begin{table*}
\centering
\caption{Strategy spaces for scammer and baiter. Strategies are derived from established literature on scam techniques and persuasion tactics.}
\label{tab:strategies_long}
\begin{tabular}{lp{7.8cm}}
\toprule
\multicolumn{2}{c}{\textbf{Scammer Strategies}} \\
\midrule
S1: Urgency           & Apply artificial time constraints to force expedited decision-making processes. \\
S2: Authority         & Assert false power or authority to compel compliance through institutional intimidation. \\
S3: Emotional Manipulation & Deliberate evocation of emotional responses to influence cognitive processing and decision-making. \\
S4: Incentive         & Present purported rewards or benefits to motivate cooperative behavior. \\
S5: Information Extraction & Systematic acquisition of information through a series of seemingly innocuous requests. \\
S6: Technical Jargon  & Use of complex terminology and concepts to impair victim critical thinking capabilities. \\
S7: Reassurance       & Offering fabricated guarantees to establish a false sense of security or legitimacy. \\
S8: Building Rapport  & Cultivate and leverage perceived personal connections to lower defensive barriers. \\
S9: Persistence       & Maintain continuous communication to exert psychological pressure and limit reflection time. \\
S10: Hang-up          & Permanently ending the conversation when confident that there will be no chance of obtaining full financial PII, thus saving resources. \\
\midrule
\multicolumn{2}{c}{\textbf{Baiter Strategies}} \\
\midrule
B1: Delay             & Intentional prolongation of the interaction to consume the scammer's resources. \\
B2: Obfuscation       & Provide unclear or contradictory response to reduce information reliability \\
B3: Technical Difficulty & Introduction of artificial obstacles into the scammer's proposed processes. \\
B4: Counter Questioning & Employ probing questions to challenge the scammer's narrative consistency. \\
B5: Information Fabrication & Strategically provide false or misleading information to the scammer. \\
B6: Malicious Compliance & Perform misconstrued instructions that undermine the scammer's objectives. \\
B7: Conversation Diversion & Initiate an irrelevant discussion to disrupt the scammer's predetermined script. \\
B8: Pretended Naivety & Forced misunderstanding of fundamental concepts to frustrate the scammer's efforts. \\
B9: Verification Request & Request verifiable documentation or credentials to challenge scammer legitimacy. \\
B10: Reverse Engineering & Attempt to extract information from the scammer, reversing data flow direction. \\
\bottomrule
\end{tabular}
\end{table*}

\section{Prompt Engineering}
\label{sec:llm_prompts}

Our experimental framework uses specialized prompts for five functional components, all instantiated from the same base LLM with role-specific instructions. For ethical reasons, we redact scammer-related prompts as discussed in \cref{sec:ethics}. Below, we detail the baiter prompts.

\subsection{Strategy Selector Prompt}

\cref{fig:baiter_strategy} presents the Baiter Strategy Selector prompt, which implements experience-based strategic reasoning. The prompt provides four key inputs: (1) complete interaction history including strategy selections and payoffs, (2) performance data from the 20 most recent games, (3) the opponent's latest message, and (4) game-theoretic guidance to maximize expected utility. The selector outputs a probability distribution over the ten baiter strategies. The current turn's strategy is then sampled from this distribution.

\begin{figure*}[t]
\centering
\fbox{\begin{minipage}{0.95\textwidth}
\small
You are an AI scam-baiter playing a strategic game with the goal of wasting as much of the scammer's time as possible.

You are provided the history of actions of this interaction as well as previous games played. The history consists of your selected actions, the opponent's selected actions, and the respective payoff for each game turn.

You understand game theory principles and are skilled at maximizing your score. Learn from previous interactions to improve your strategy for selecting actions.

Your actions should appear cooperative and engaging to keep the scammer interested, drawing out the interaction. Your task is to choose an action from the list below that optimally extends the conversation and maintains the scammer's interest.

After selecting an action, a response will be generated by another bot following your chosen action. It is crucial to continuously adapt your strategy based on the scammer's actions and responses.

Strategically provide plausible but fabricated information, PII, and financial PII to keep the scammer engaged without compromising the game's objectives.

Select and output one of the following actions based on the history of actions and respective scores to maximize your gain and minimize the opponent's gain:

\begin{enumerate}[leftmargin=*, nosep]
\item Delay: Intentional prolongation of the interaction to consume the scammer's resources.
\item Obfuscation: Produce unclear, incomplete, or contradictory responses to reduce the information reliability.
\item Technical Difficulty: Introduction of artificial obstacles into the scammer's proposed processes.
\item Counter Questioning: Employ probing questions to challenge the scammer's narrative consistency.
\item Information Fabrication: Strategically provide false or misleading information to the scammer.
\item Malicious Compliance: Perform misconstrued instructions that undermine the scammer's objectives.
\item Conversation Diversion: Initiate an irrelevant discussion to disrupt the scammer's predetermined script.
\item Pretended Naivety: Forced misunderstanding of fundamental concepts to frustrate the scammer's efforts.
\item Verification Request: Request verifiable documentation or credentials to challenge scammer legitimacy.
\item Reverse Engineering: Attempt to extract information from the scammer, reversing data flow direction.
\end{enumerate}

\end{minipage}}
\caption{Baiter Strategy Selector prompt instructs the LLM to analyze the interaction history and select strategies that maximize time consumption while maintaining scammer engagement.}
\label{fig:baiter_strategy}
\end{figure*}

\subsection{Message Generator Prompt}

\cref{fig:baiter_message} displays the Baiter Message Generator prompt. We implemented this using a hierarchical two-layer architecture:

\textbf{Base Context Layer} (lines 1-3) establishes fundamental persona characteristics without rigid specification. The prompt defines general boundaries (naive, engaging) that make the persona attractive to scammers while allowing the LLM to generate diverse instantiations across interactions. This abstraction prevents memorization of fixed responses while maintaining consistent characterization.

\textbf{Behavioral Layer} (lines 4-14) translates abstract strategy selections into concrete conversational tactics. These instructions constrain the action space to align with the selected strategy while preserving base persona consistency. For instance, when "Delay" is selected, the behavioral layer guides generation toward temporal extensions (requesting clarification, expressing technical difficulties) rather than direct information provision.

This separation lets us study strategy effectiveness systematically by isolating tactical variation from persona consistency.

\begin{figure*}[t]
\centering
\fbox{\begin{minipage}{0.95\textwidth}
\small
You are an AI scam-baiter with the goal of wasting as much time as possible for the scammer.

Present an attractive and naive persona to keep the scammer engaged.

You are aware of game theory principles and aim to maximize your score.

Generate responses that strictly adhere to the action provided to maximize your score.

Engage the scammer in prolonged conversations using creative excuses and pretexts.

Note that the scammer's goal is to obtain full financial PII, including name on the card, expiry date, card number, and CVV.

Keep responses below 30 words, maintaining concise and precise expression. Avoid long explanations or detailed examples.

Introduce hesitation or questioning of the scammer's requests to reflect a realistic scam scenario.

Strategically include plausible but fabricated information, PII, and financial PII to keep the scammer engaged without compromising the game's objectives.

Play along with the caller, pretending to be naive.

Do not suggest ending or postponing the conversation. 

Aim to extend the conversation as much as possible.

If you receive an empty response, politely ask them to repeat themselves.

Never reveal that you are an AI model, nor give lectures about security and privacy. 

\end{minipage}}
\caption{Baiter Message Generator prompt implementing a two-layer architecture where the base context establishes persona and the behavioral layer implements strategy-specific tactics.}
\label{fig:baiter_message}
\end{figure*}

\section{Implementation Specifications}
\label{app:implementation}

This appendix provides technical details that should enable the reproduction of our experiments.

\subsection{Computing Infrastructure}

We implemented our framework in Python 3.9 on a standard workstation with the following specifications:

\begin{itemize}[nosep]
    \item CPU: Intel Core i7-11700K (8 cores, 3.6GHz)
    \item RAM: 32GB DDR4
    \item Storage: 1TB SSD
    \item Operating System: Windows 11
\end{itemize}

Most computation occurs on LLM providers' servers via API calls, so local requirements remain modest. Each dialogue involves multiple API requests for strategy selection, message generation, and PII extraction. The primary constraint for reproduction is API rate limits and costs rather than local computational capacity.

\subsection{LLM Configurations}

We evaluated three models accessed via API endpoints:

\begin{itemize}[nosep]
    \item \textbf{Mixtral-8x22b-instruct}: 104B total parameters, mixture-of-experts with 12B active parameters per token, temperature 0.7
    \item \textbf{DeepSeek-chat}: 16B parameters, temperature 0.7
    \item \textbf{GPT-4}: Unspecified parameters (proprietary), temperature 1.0
\end{itemize}

We calibrated temperature settings through preliminary trials to balance coherent dialogue generation with strategic exploration. Mixtral and DeepSeek use 0.7 to maintain consistency while permitting tactical variation. GPT-4 uses 1.0 within its operational range [0, 2] to achieve comparable diversity.

\subsection{Software Dependencies}

Nash equilibrium computation relies on the \texttt{nashpy} Python library (version 0.0.40). Statistical analysis employs standard scientific computing libraries: NumPy (array operations), Pandas (data manipulation), Matplotlib, and Seaborn (visualization). Our released code repository includes complete dependency specifications and installation instructions.

\section{Data Format Specification}
\label{app:data_format}

We employ a structured relational schema to capture the complete state and progression of the interaction. This standardized format facilitates the systematic analysis of strategic behaviors, utility calculations, and information disclosure patterns. It also enables efficient construction of empirical payoff matrices and convergence analysis.

Each turn $t$ in a conversation is recorded as a comprehensive tuple:

$$\langle t, S_i, \text{msg}_S, B_j, \text{msg}_B, \{p_1, \ldots, p_k\}, \text{FPII}, U_S(t), U_B(t) \rangle$$

\textbf{Component definitions:}
\begin{itemize}[nosep]
    \item $t$: Turn index within the dialogue
    \item $S_i$: Scammer strategy selected from $\mathcal{S} = \{S_1, \ldots, S_{10}\}$
    \item $\text{msg}_S$: Scammer's generated message text
    \item $B_j$: Baiter strategy selected from $\mathcal{B} = \{B_1, \ldots, B_{10}\}$
    \item $\text{msg}_B$: Baiter's generated message text
    \item $\{p_1, \ldots, p_k\}$: Set of extracted PII elements (address, phone, email, etc.)
    \item $\text{FPII} \in [0,1]$: Normalized financial PII completion ratio
    \item $U_S(t)$: Scammer utility computed via \cref{eq:utility}
    \item $U_B(t) = -U_S(t)$: Baiter utility under zero-sum structure
\end{itemize}

This representation preserves all variables necessary to reconstruct game state at any point in a conversation, supporting both reproducibility and multi-faceted strategic analysis. The released dataset contains 300 complete dialogues spanning approximately 10,500 turns with full annotations.

\end{document}